\documentclass[preprint,epsfig,aps,showpacs]{revtex4}
\usepackage{amsmath}
\usepackage{graphicx}

\newcommand{\f}{\begin{equation}}
\newcommand{\ff}{\end{equation}}
\newcommand{\fa}{\begin{eqnarray}}
\newcommand{\ffa}{\end{eqnarray}}

\begin{document}
\title{Late-time phantom universe in ${\bf SO(1,1)}$ dark energy model with exponential potential}
\author{Yi-Huan Wei${ }^{1,2}$}
\affiliation{%
${ }^1$ Department of Physics, Bohai University, Jinzhou 121000,
 Liaoning, China}
\affiliation{%
$ { }^2$ Institute of Theoretical Physics, Chinese Academy of
Science, Beijing 100080, China}
\begin{abstract}
~~We discuss the late-time property of universe and phantom field
in the $SO(1,1)$ dark energy model for the potential
$V=V_0e^{-\beta\Phi^\alpha}$ with $\alpha$ and $\beta$ two
positive constants. We assume in advance some conditions satisfied
by the late-time field to simplify equations, which are confirmed
to be correct from the eventual results. For $\alpha<2$, the filed
falls exponentially off and the phantom equation of state rapidly
approaches $-1$. When $\alpha=2$, the kinetic energy $\rho_k$ and
the coupling energy $\rho_c$ become comparable but there is always
$\rho_k<-\rho_c$ so that the phantom property of field proceeds to
hold. The analysis on the perturbation to the late-time field
$\Phi$ illustrates the square effective mass of the perturbation
field is always positive and thus the phantom is stable. The
universe considered currently may evade the future sudden
singularity and will evolve to de Sitter expansion phase.

\end{abstract}

\maketitle

The accelerated expansion of the universe \cite{Riess} indicates
the existence of energy component with big negative pressure, dark
energy \cite{Quin,K,T,C,Boyle}, which weakly interacts with
ordinary matter. The current observations \cite{Rie,Wang} hint
that dark energy may likely possess a super-negative pressure,
$w_X<-1$, i.e., phantom
\cite{Caldwell,Caldwel,Mel,P,Diaz,Sami,Abdalla,Lue,Cal}. There are
at least the two classes of phantom models, one characterized by
the equation of state $p_X=w_X\rho_X$, others satisfying some
different equation of state such as Chaplygin gas \cite{C}. For
the first class, a special case is that $w_X$ is a constant, in
which a Big Rip occurs \cite{Caldwell,McInnes}. Phantom violates
the weak energy condition, but it may be stable providing that the
second derivative of potential with respect to filed is negative
\cite{Mel,Cal}. Phantom particle may be allowed by conventional
kinematics, ordinary particles can decay into heavier particles
plus phantoms \cite{Mel}. Compared with the age of the universe,
if phantom particle's life is long enough, then it can also be
thought to be stable.

Either a phantom universe may evolve to or avoid the future
singularity depends on the used model. For the generalized
Chaplygin gas as phantom, the universe may avoid Big Rip
\cite{Gonzalez}. Quantum phantom effects may evade the occurrence
of singularity since the universe may end up in de Sitter phase
before scale factor blows up \cite{Nojiri}. In a higher-derivative
scalar-tensor theory, the universe can admit an effective phantom
description with a transient acceleration phase \cite{Elizalde}.
For a overall phantom expansion phase with $p_X=w_X\rho_X$, what
about this case? For a constant $w_X$, there is the scale factor
$a\sim [-w_Xt_m+\gamma_X t]^{2/3\gamma_X}$ with
$\gamma_X=1+w_{X}$, which implies a Big Rip will occur in a finite
time. Here, we consider the case of time-dependent $w_X$. Treating
it as the approximate expression to $a$ for a changing $w_X$, then
we conjecture for $\gamma_X=-Wt^{-n}$ with $W>0$ and $n>1$ two
constants, $|\gamma_X t|=Wt^{1-n}$ decreases with time so that the
universe, as pointed out in \cite{Wei}, may escape future Big Rip.
For an exponentially evolving $\gamma_X\sim-e^{\sigma t}$ with
$\sigma$ a negative constant, $|\gamma_X t|$ may diminish with
time more rapidly than the power-law one mentioned above for late
time, so one can conjecture in this case the universe can also
avoid future sudden singularity.

The purpose of this paper is to discuss the problem of the
avoidance of Big Rip in the context of $SO(1,1)$ model. We find
this model can give a distinctive description for the phantom
field and the universe due to existence of the symmetry. We
consider the case of the potential
$V(\Phi)=V_0e^{-\beta\Phi^{\alpha}}$ with $\alpha\leq2$ and
analyze the properties of the late-time field the universe. For
this exponential potential, the late-time field $\Phi$
exponentially decreases to zero, the scale factor $a$
exponentially increases and the equation of state tends to $-1$. A
simple analysis on the perturbation of field shows the late-time
phantom is stable and the universe can evade a Big Rip.

Before analyzing the late-time evolution of phantom field and
universe, we give a simple discussion of $SO(1,1)$ model. This
model contains two field components, $\phi_1$ and $\phi_2$, or
equivalently, $\Phi$ and $\theta$. For late time, the matter
density is much smaller than phantom energy density and equations
(6)-(11) given in Ref. \cite{Wei} reduce to
\begin{eqnarray}
H^2=(\frac{\dot{a}}{a})^2=\frac{8\pi G}{3}\rho_\Phi, \quad
\frac{\ddot{a}}{a}=-\frac{4\pi G}{3}(\rho_\Phi+3p_\Phi),
\label{eq1}
\end{eqnarray}
\begin{eqnarray}
\ddot{\Phi}+3H\dot{\Phi}+\dot{\theta}^2\Phi+V'(\Phi)=0,
\label{eq2}
\end{eqnarray}
\begin{eqnarray}
\dot{\theta}=\frac{c}{a^3\Phi^2}, \label{eq3}
\end{eqnarray}
\begin{eqnarray}
\rho_\Phi=\frac12(\dot{\Phi}^2-\Phi^{2}\dot{\theta}^{2})+V(\Phi),
\quad
p_\Phi=\frac12(\dot{\Phi}^2-\Phi^{2}\dot{\theta}^{2})-V(\Phi).
\label{eq4}
\end{eqnarray}
where $c$ is a constant.

Noting that
$\ddot{\theta}=-(2\frac{\dot{\Phi}}{\Phi}+3H)\dot{\theta}$ and
$\dot{\rho}_\Phi=\dot{\Phi}\ddot{\Phi}+\Phi\dot{\Phi}\dot{\theta}^2+3H\Phi^2\dot{\theta}^2+V'\dot{\Phi}$,
then one can find (\ref{eq2}) will yields
\begin{eqnarray}
\dot{\rho}_\Phi+3H(\rho_\Phi+p_\Phi)=0. \label{eq5}
\end{eqnarray}
Equation (\ref{eq5}) implies that (\ref{eq2}) plays practically
the role of the conserved equation in $SO(1,1)$ dark energy model.

The exponential potential has greatly been studied in
quintessence, phantom and other models \cite{Steinhardt}. For the
Gaussian potential, $V=V_0e^{-\Phi^{2}/\sigma^2}$, a numerical
analysis shows that the phantom universe will evolve regularly
\cite{Mel}. Here, for $SO(1,1)$ model we consider the following
exponential potential
\begin{eqnarray}
V=V_0e^{-\beta\Phi^{\alpha}},  \label{eq6}
\end{eqnarray}
where $\alpha$ and $\beta$ are two positive constants. For
potential (\ref{eq6}), there are two possible cases for the
late-time evolution of field, $\Phi$ increases or decreases with
time. We suppose the field should decrease with time and assume
the field satisfies the following condition in advance
\begin{eqnarray}
\dot{\theta}^2\Phi^2\gg\dot{\Phi}^2, \quad
\dot{\theta}^2\Phi\gg\ddot{\Phi}, \quad
\dot{\theta}^2\Phi\gg3H\dot{\Phi}. \label{eq7}
\end{eqnarray}
Letting
\begin{eqnarray}
\ddot{\Phi}+3H\dot{\Phi}=\eta_1\dot{\theta}^2\Phi, \label{eq8}
\end{eqnarray}
where $\eta_1=\eta_1(t)$ is a small quantity and putting
(\ref{eq3}) and $V'=-\alpha \beta\Phi^{\alpha-1}V$ in (\ref{eq2}),
then we have
\begin{eqnarray}
\frac{c^2(1+\eta_1)}{a^6\Phi^3}-\alpha \beta\Phi^{\alpha-1}V=0,
\label{eq9}
\end{eqnarray}
which yields
\begin{eqnarray}
a=A^{\frac{1}{6}}\Phi^{\frac{-\alpha-2}{6}}e^{\beta\Phi^{\alpha}/6},
\quad A=\frac{c^2(1+\eta_1)}{\alpha\beta V_0}. \label{eq10}
\end{eqnarray}
Eq. (\ref{eq10}) shows the simple relation between $a$ and $\Phi$,
from which the Hubble parameter may be expressed in terms of
$\Phi$ and $\eta_1$ as
\begin{eqnarray}
H=\frac{\dot{a}}{a}=\frac{1}{6}[(1+\eta_1)^{-1}\dot{\eta}_1-(\alpha+2)\Phi^{-1}\dot{\Phi}
+\alpha\beta\Phi^{\alpha-1}\dot{\Phi}], \label{eq11}
\end{eqnarray}
in which the first term on the right-hand side is small quantity.

Defining the kinetic energy and the coupling energy as
\begin{eqnarray}
\rho_k=\frac{1}{2}\dot{\Phi}^2, \quad \rho_c=-c^2\Phi^{-2}a^{-6},
\label{eq12}
\end{eqnarray}
and letting
\begin{eqnarray}
\rho_k=-\eta_2\rho_c, \label{eq13}
\end{eqnarray}
where $\eta_2=\eta_2(t)$ is also a small quantity, then the total
energy density is
\begin{eqnarray}
\rho_\Phi=\rho_k+\rho_c+V=[1-\frac{1-\eta_2}{1+\eta_1}\alpha
\beta\Phi^{\alpha}]V, \label{eq14}
\end{eqnarray}
where $\Phi\leq[\frac{1-\eta_2}{1+\eta_1}\alpha
\beta]^{-1/\alpha}\simeq(\alpha \beta)^{-1/\alpha}$ which
guarantees positivity of $\rho_\Phi$, noting that $\eta_1$ and
$\eta_2$ are two small quantities. From Eqs. (\ref{eq1}) and
(\ref{eq11}), dropping the first term on the right-hand side of
the latter we have
\begin{eqnarray}
\frac{1}{6}[\alpha\beta\Phi^{\alpha-1}-(\alpha+2)\Phi^{-1}]\dot{\Phi}
\simeq\mu_P^{-1}\rho_\Phi^{1/2}, \label{eq15}
\end{eqnarray}
where $\mu_P^2=3M_P^2$ with $M_P=1/\sqrt{8\pi G}$ is the reduced
Planck mass. As the phantom solution of (\ref{eq15}), $\rho_\Phi$
must increase with time, this implies $\Phi$ should diminish with
time. Assuming that $\Phi\ll (\alpha\beta)^{1/\alpha}$ and
$\Phi\ll (\beta)^{1/\alpha}$ for $t\gg 1$, then there are
$e^{-\frac{\beta}{2}\Phi^{\alpha}}\simeq1$ and $\rho_\Phi\simeq
V_0$; neglecting the first term on the left-hand, approximately to
the lowest order then (\ref{eq15}) reduces to
\begin{eqnarray}
-\frac{\alpha+2}{6}\Phi^{-1}\dot{\Phi}\simeq\mu_P^{-1}\sqrt{V_0}.
\label{eq16}
\end{eqnarray}
Solving (\ref{eq16}) and putting $\Phi$ in (\ref{eq10}), then we
obtain
\begin{eqnarray}
\Phi=\Phi_0e^{-\frac{6\sqrt{V_0}}{\mu_P(\alpha+2)}t}, \quad
a=a_0e^{\frac{
\sqrt{V_0}}{\mu_P}t}e^{\frac{\beta}{6}\Phi^{\alpha}},
\label{eq17}
\end{eqnarray}
where $\Phi_0$ is a constant and
$a_0=A^{\frac{1}{6}}\Phi_0^{-\frac{\alpha+2}{6}}$. Eq.
(\ref{eq17}) shows $\Phi$ fall exponentially off and the Hubble
parameter approaches $H_m=\frac{\sqrt{V_0}}{\mu_P}$.

Expanding
$\sqrt{1-\frac{1-\eta_2}{1+\eta_1}\alpha\beta\Phi^{\alpha}}\simeq
1-\frac{1-\eta_2}{2(1+\eta_1)}\alpha\beta\Phi^{\alpha}$,
$e^{-\frac{\beta}{2}\Phi^{\alpha}}\simeq
1-\frac{\beta}{2}\Phi^{\alpha}$ and letting $\chi=
(\alpha+2)[\frac{1-\eta_2}{1+\eta_1}\alpha+1]-2\alpha$ which is a
slowly-changing function and thus is treated as a constant here,
then Eq. (\ref{eq15}) gives
\begin{eqnarray}
-\frac{1}{12}[2(\alpha+2)\Phi^{-1}-\chi\beta\Phi^{\alpha-1}]\dot{\Phi}
\simeq H_m, \label{eq18}
\end{eqnarray}
which yields
\begin{eqnarray}
\Phi
e^{\frac{\chi\beta}{2\alpha(\alpha+2)}\Phi^{\alpha}}=\Phi_0e^{-\frac{6}{\alpha+2}H_mt},
\label{eq19}
\end{eqnarray}
or approximately
\begin{eqnarray}
\Phi\simeq\Phi_0e^{-\frac{6}{\alpha+2}H_mt}-\frac{\chi\beta}{2\alpha(\alpha+2)}\Phi_0^{\alpha+1}
e^{-\frac{6(\alpha+1)}{\alpha+2}H_mt}. \label{eq20}
\end{eqnarray}
Putting (\ref{eq20}) in (\ref{eq6}) and (\ref{eq10}) yields the
late-time scale factor and the potential
\begin{eqnarray}
a\simeq a_0
[1+\frac{(2\alpha+\chi)\beta\Phi_0^{\alpha}}{12\alpha}e^{-\frac{6\alpha}{\alpha+2}H_mt}]e^{H_mt},
\label{eq21}
\end{eqnarray}
\begin{eqnarray}
V\simeq V_0e^{-\beta\Phi_0^\alpha
e^{-\frac{6\alpha}{\alpha+2}H_mt}}\simeq V_0(1-\beta\Phi_0^\alpha
e^{-\frac{6\alpha}{\alpha+2}H_mt}). \label{eq22}
\end{eqnarray}

In deriving (\ref{eq20})-(\ref{eq22}), the condition (\ref{eq7})
has been assumed. A check for it must be done. Holding
$\eta_1,\eta_2$ but neglecting their derivatives with respect to
time, then we obtain
\begin{eqnarray}
\rho_k=\frac{18\Phi_0^2}{(\alpha+2)^2}H_m^2e^{-\frac{12}{\alpha+2}H_mt},
\quad \rho_c=-\frac{\alpha\beta}{1+\eta_1}\Phi_0^\alpha
V_0e^{-\frac{6\alpha}{\alpha+2}H_mt}, \label{eq23a}
\end{eqnarray}
\begin{eqnarray}
\dot{\theta}^2\Phi=
\frac{\alpha\beta}{1+\eta_1}\Phi_0^{\alpha-1}V_0e^{-
\frac{6(\alpha-1)}{\alpha+2}H_mt}, \label{eq23b}
\end{eqnarray}
\begin{eqnarray}
\ddot{\Phi}=
\frac{36}{(\alpha+2)^2}H_m^2\Phi_0e^{-\frac{6}{\alpha+2}H_m t},
\quad 3H\dot{\Phi}=
-\frac{18}{\alpha+2}H_m^2\Phi_0e^{-\frac{6}{\alpha+2}H_mt},
\label{eq23c}
\end{eqnarray}
which clearly satisfy (\ref{eq7}) for $\alpha<2$. From
(\ref{eq23a})-(\ref{eq23c}), for $\alpha<2$ the two small
quantities $\eta_1$ and $\eta_2$ are determined as
\begin{eqnarray}
\eta_1\simeq-\frac{6\Phi_0^{2-\alpha}}{(\alpha+2)^2\beta
M_P^2}e^{\frac{6(\alpha-2)}{\alpha+2}H_mt}, \quad
\eta_2\simeq\frac{6\Phi_0^{2-\alpha}}{\alpha(\alpha+2)^2 \beta
M_P^2}e^{\frac{6(\alpha-2)}{\alpha+2}H_mt}, \quad \label{eq26}
\end{eqnarray}
which implies the fact that $\chi$ is indeed a slowly-changing
quantity. In deriving (\ref{eq26}), $H_m^2=V_0/\mu_P^2$ and
$\mu_P^2=3M_P^2$ have been used. From (\ref{eq22}) and
(\ref{eq23a}), barotropic index
$\gamma_\Phi=1+w_\Phi=\frac{2(\rho_k+\rho_c)}{\rho_k+\rho_c+V}$ is
given as
\begin{eqnarray}
\gamma_\Phi=-\frac{2\alpha\beta}{1+\eta_1} \Phi_0^\alpha
e^{-\frac{6\alpha}{\alpha+2}H_mt}+\frac{12\Phi_0^2}{(\alpha+2)^2M_P^2}e^{-\frac{12}{\alpha+2}H_mt},
\label{eq27}
\end{eqnarray}
which shows $\gamma_\Phi$ decreases exponentially to zero and thus
implies the universe will approach the de Sitter phase with $H_m$
the expansion rate.

For $\alpha=2$, the condition (\ref{eq7}) is no longer true.
However, the results (\ref{eq20})-(\ref{eq23c}) and (\ref{eq27})
are still correct, this may be seen from the process of obtaining
them and the fact that $\eta_1,\eta_2$ are two finite constants.
For $\alpha=2$, (\ref{eq22})-(\ref{eq23c}) gives rise to
\begin{eqnarray}
\rho_k=\frac{3\Phi_0^2V_0}{8M_P^2}e^{-3H_mt}, \quad
\rho_c=-\frac{2\beta\Phi_0^2V_0}{1+\eta_1} e^{-3H_mt}, \quad
V=V_0(1-\beta\Phi_0^2 e^{-3H_mt}), \label{eq28}
\end{eqnarray}
which shows $\rho_k$ and $\rho_c$ fall off according to the same
exponential law, and
\begin{eqnarray}
\dot{\theta}^2\Phi= \frac{2\beta\Phi_0V_0}{1+\eta_1}e^{-3H_mt/2},
\quad \ddot{\Phi}= \frac{3\Phi_0V_0}{4M_P^2}e^{-3H_m t/2}, \quad
3H\dot{\Phi}= -\frac{3\Phi_0V_0}{2M_P^2}e^{-3H_m t/2}.
\label{eq29}
\end{eqnarray}

Using $\ddot{\Phi}+3H\dot{\Phi}=\eta_1\dot{\theta}^2\Phi$ and
$\rho_k=-\eta_2\rho_c$, then from (\ref{eq28}) and (\ref{eq29}) it
follows
\begin{eqnarray}
\frac{1+\eta_1}{2\beta\eta_1}=-\frac{4}{3}M_P^2, \quad
\frac{1+\eta_1}{2\beta\eta_2}=\frac{8}{3}M_P^2, \label{eq30}
\end{eqnarray}
which yield
\begin{eqnarray}
\eta_1=-\frac{3}{3+8\beta M_P^2}, \quad \eta_2=\frac{3}{2(3+8\beta
M_P^2)}. \label{eq31}
\end{eqnarray}
Importantly, Eq. (\ref{eq31}) tells $\eta_2<\frac{1}{2}$, this
means that for the exponential potential (\ref{eq6}) with
$\alpha\leq2$, the model always has the late-time phantom
property.

Here, we give a simple analysis on the stability. For simplicity,
assuming that the fluctuations of the field $\theta$ and the
background may be neglected, taking the perturbation to the field
$\Phi$ as $\delta\Phi=\sum\delta\Phi_k(t)e^{i\vec{k}\cdot
\vec{x}}$, where k denotes the mode with energy $k$, then the
evolution of perturbation field $\delta\Phi$ is determined by
\begin{eqnarray}
\ddot{\delta\Phi_k}(t)+3H\dot{\delta\Phi_k}(t)+\big(V''+\dot\Theta^2+k^2)
{\delta\Phi_k}(t)=0 \label{eq32},
\end{eqnarray}
where $V''$ denotes the second order derivative with respect to
$\Phi$. From Eqs. (\ref{eq2}), (\ref{eq6}) and (\ref{eq8}) it
follows that
$V''=[\alpha\beta(1-\alpha)\Phi^{\alpha-2}+\alpha^2\beta^2\Phi^{2\alpha-2}]V$
and $\dot{\Theta}^2=\frac{\alpha\beta}{1+\eta_1}\Phi^{\alpha-2}V$.
The effective mass of $\delta\Phi$ is given by
\begin{eqnarray}
m_{eff}^2=V''+\dot\Theta^2+k^2=\alpha\beta(1+\frac{1}{1+\eta_1}-\alpha)\Phi^{\alpha-2}V+
\alpha^2\beta^2\Phi^{2\alpha-2}V+k^2 \label{eq33}.
\end{eqnarray}
Eq. (\ref{eq33}) shows the effective mass of the perturbation
field differs from that given in \cite{Cal,Mel}, the obvious
distinctions between them are the positive sign of $V''$ and
existence of the positive term $\dot\Theta^2$ which can strength
the stability of the filed. Noting that $V\simeq V_0$ for late
time and $\eta_1$ and $\Phi$ are small quantity for $\alpha<2$,
then there approximately is
$m_{eff}=[\alpha\beta(2-\alpha)V_0\Phi^{\alpha-2}]^{\frac{1}{2}}$
which depends on time and is large. For $\alpha=2$, inserting
$\eta_1=-\frac{3}{3+8\beta M_P^2}$ in (\ref{eq33}) leads to
$m_{eff}=[\frac{3V_0}{4M_P^2}+k^2]^{\frac{1}{2}}$. It is clear
that for $\alpha\leq2$, $m_{eff}$ is always real for both $V''<0$
and $V''>0$ so that the amplitude of the perturbation may always
be suppressed on both small and large scale and thus the phantom
field $\Phi$ in the potential (\ref{eq6}) with $\alpha\leq2$ is
stable for late time.

The $SO(1,1)$ model may be obtained by the substitution
$Q\rightarrow ic$ from the $U(1)$ model, a generalization to
quintessence \cite{Boyle}, and it can be considered as either the
quintessence or phantom model. Compared with the phantom model
given from quintessence by $\phi\rightarrow i\phi$, the $SO(1,1)$
phantom model contains an extra negative coupling energy, so it
certainly possesses some different properties and may be expected
to give a distinct description for phantom. We have studied the
late-time evolution properties of the universe and the field for
the exponential potential. In this model, the late-time universe
and field appears some distinctive features. Compared with the
coupling energy, the kinetic energy of the late-time field is a
small quantity for $\alpha<2$. Unlike the phantom with negative
kinetic energy, it is the case that negative coupling energy
exceeds positive kinetic energy results in the super-negative
pressure and therefore the behavior of field is quite different
from that in the former phantom model. In the former model,
stability on the large scale requires $V''<0$, however, in the
latter no constraint on the potential from stability is needed.
The universe driven by the $SO(1,1)$ phantom with the exponential
potential can avoid a future Big Rip and will evolve to the
approximate de Sitter expansion phase.

\vskip 3.9cm {\bf Acknowledgement:} {This work are supported by
ITP Post-Doctor Project 22B580, Chinese Academy of Science of
China, National Nature Science Foundation of China and Liaoning
Province Educational Committee Research Project.}

\end{document}